\documentclass[12pt,nofootinbib,showpacs]{revtex4-1}

\usepackage{amsmath}
\usepackage{graphicx}

\begin{document}

\newcommand\pdag{{\vphantom\dagger}}
\newcommand\ket[1]{\left| #1 \right\protect\rangle}

\title{Measurement-induced entanglement of two transmon qubits by a single photon}
\author{Christoph Ohm and Fabian Hassler}
\affiliation{JARA Institute for Quantum Information, RWTH Aachen University, 52056 Aachen, Germany}

\date{October 2016}

\begin{abstract} 
On-demand creation of entanglement between distant qubits is a necessary
ingredient for distributed quantum computation. We propose an entanglement
scheme that allows for single-shot deterministic entanglement creation by
detecting a single photon passing through a Mach-Zehnder interferometer with
one transmon qubit in each arm. The entanglement production essentially relies
on the fact that superconducting microwave structures allow to achieve strong
coupling between the qubit and the photon. By detecting the photon via a
photon counter, a parity measurement is implemented and the wave function of
the two qubits is projected onto a maximally entangled state. Most
importantly, the entanglement generation is heralded such that  our protocol
is not susceptible to photon loss due to the indivisible nature of single
photons.

\noindent \textit{Keywords}: superconducting qubits, entanglement, single photons

\noindent \textit{Email:} \texttt{hassler@physik.rwth-aachen.de}
\end{abstract}

\pacs{
03.67.Bg 
42.50.Dv 
42.50.Pq 
85.25.--j 
}

\maketitle 

Entanglement is one of the most characteristic features distinguishing quantum
mechanics from classical mechanics \cite{Horodecki:2009fk} and its paradoxical
predictions have challenged generations of physicists, see e.g.
\cite{Einstein:1935aa}. Quantum information theory aims to exploit
entanglement as a resource for protocols guaranteeing (secure) quantum
communication over macroscopic distances, used in quantum teleportation
\cite{Bennett:1993lr}, quantum key distribution \cite{Ekert:1991fk}, and
distributed quantum computation \cite{Nielsen:2010fk}. 

The pioneering works of \cite{Yurke:1992uq,Yurke:1992lr} and
\cite{Zfiukowski:1993qy} have shown that entanglement can not only be
transferred via direct interactions but also by performing a measurement such
that the wave function is projected onto an entangled state. This method,
known as  measurement-based entanglement, is based on the indistinguishability
of the quantum states compatible with the measurement outcome and constitutes
one of the key ingredients used for quantum repeaters \cite{Briegel:1998fk}.
Moreover, the genesis of entanglement by performing a measurement has been
first proposed for atoms in a quantum optical framework
\cite{Cabrillo:1999kx,Plenio:1999ul}. Since then, measurement-induced
entanglement of remote quantum systems has been experimentally demonstrated
for diverse atomic setups \cite{Chou:2005vn, Moehring:2007fr, Hofmann:2012lr}
as well as for solid state qubit devices such as nitrogen vacancy centres
\cite{Bernien:2013fk} and superconducting qubits \cite{Roch:2014qy,narla:16}.
\begin{figure}[t]
  \centering
  \includegraphics[width=0.6\textwidth]{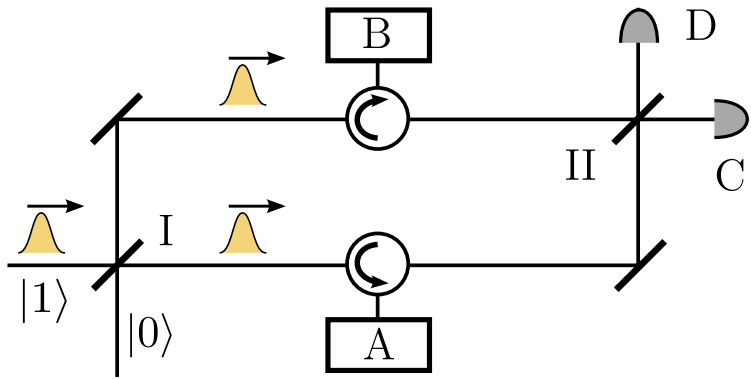} 
  \caption{Mach-Zehnder interferometer made of two 50:50 beamsplitters (I and
   II) implemented via directional couplers. Each arm of the interferometer,
   formed by a coplanar waveguide, is coupled to a microwave resonator (A and
   B) which in turn is dispersively coupled to a transmon qubit.  A
   single-photon wave-packet that is send into the MZI causes a click in one
   of the two detectors at the outputs. The projective measurement due to the
   photon counters implements a parity measurement on the transmon qubits.  As
   a result, irrespective of which photon detector (C or D) clicks, the wave
   function is projected onto a maximally entangled two-qubit state.  }
  \label{fig:setup}
\end{figure}

In the optical domain, the light-matter interaction is rather weak. To
overcome this problem most quantum optical entangling-schemes exploit the
photon polarization degree of freedom, whereas other proposals suggest to use
more challenging concepts such as NOON-states \cite{Huang:2008lr}. In
contrast, in circuit quantum electrodynamics (cQED), which deals with
superconducting qubits and their interaction to microwave radiation, it has
been possible to reach the strong coupling regime in which the light-matter
coupling is stronger than typically induced dissipation scales
\cite{Wallraff:2004fk}.  Even though microwave photons are typically
unpolarized, present cQED-implementations accomplished measurement-induced
entanglement of superconducting qubits employing coherent states, either with
the qubits placed inside a single cavity \cite{Riste:2013yq} or in separate
resonators \cite{Roch:2014qy}. Due to the lack of erasure of the which-path
information, these protocols only offer a maximal success rate of 50\%. More
recently, it has been understood that employing additionally a non-linear
element a success rate of 100\% is achievable in principle
\cite{Roy:fk,Silveri:uq}.  The quantum mechanical state projection is caused
in all these cases  by a weak continuous measurement \cite{Hutchison:2009fk,
Lalumiere:2010qy}. Because of that, entanglement emerges only rather slowly
with the wave function gradually collapsing in time which also allows for the
unwanted generation of multi-partite entanglement
\cite{Helmer:2009lr,Bishop:2009qy}.

Recently, a remote entanglement protocol for transmon qubits based on the
propagation of a pair of photons has been experimentally realised
\cite{narla:16}. In protocol, each transmon qubit emits a single photon. The
photons then interfere at a beam splitter before being measured in a microwave
photon detector. The protocol is insensitive to photon loss. However, due to
the fact that the protocol involves two photons, only a maximal success rate
of 50\% is possible.  In all the proposals presented so far, at least two
photons are needed to generate the remote entanglement.  Here, we propose a
novel scheme for entangling superconducting transmon qubits over a distance by
using a strong projective measurement of a \emph{single} microwave photon. The
photon propagates through a Mach-Zehnder interferometer (MZI) containing two
transmon qubits to be entangled. Relying on the discreteness of photonic Fock
states as well as the ability of cQED to access the strong coupling regime,
our scheme allows for the generation of remote entanglement in a deterministic
and on-demand fashion.  In particular, our protocol is insensitive to photon
loss and only sensitive to detector dark counts.  This is implemented by
conditioning the successful generation of entanglement on the detection of the
photon, see below.

Within the last decade, cQED has become a mature field of quantum engineering
technology promising integrated and scalable circuits suitable for quantum
computation. In particular, the generation of single microwave photons in
superconducting circuits \cite{Houck:2007ly} and moreover the controlled
creation of entanglement between microwave photons and transmon qubits has
been reported \cite{Eichler:2012lr}. These promising attempts towards
well-controlled microwave photonics encouraged us to suggest an entanglement
protocol taking advantage of the high efficiencies in cQED combined with
single-shot (projective) measurements. For a long time, the efficient
detection of single microwave photons has been an experimental challenge
\cite{Bozyigit:2011ve}. However, rapid experimental progress in detecting
single flying microwave photons \cite{Chen:2011,fan:14,Inomata:2016,narla:16}
combined with a multitude of theoretical proposal
\cite{Romero:2009fk,Silva:2010ul,Govia:2012,Govia:2014} give us confidence
that the remaining challenges will be mitigated in the near future. For
example, capturing of a single flying photon by a transmon qubit has been
recently reported with a detection efficiency of 90\% \cite{narla:16}. 

The outline of the paper is as follows. First, we will explain the generation
of entanglement by introducing the interferometric apparatus and discuss the
procedure under ideal conditions. Subsequently, a more general analysis
clarifies the conditions under which maximal entanglement is achievable.
Specifically, we will discuss the case of entanglement generated by a
Lorentzian-shaped single-photon wave-packet as well as the entanglement in
case of non-identical cavities.

At the first beamsplitter, an incoming photon is split into two partial waves
$\ket{\psi_\text{ph}} = ( \ket{1_\text{A},0_\text{B}} +
\ket{0_\text{A},1_\text{B}} )/ \sqrt{2}$, each traversing one arm (A or B) of
the interferometer. After having passed the second beamsplitter, the partial
waves are recombined and the photon escaped into one of the two output
channels C and D, see figure~\ref{fig:setup}. If both partial waves acquire a
relative phase difference of $\varphi=0$ while passing the MZI, i.e., if both
partial waves accumulate exactly the same phase in both arms A and B, the
photon is transmitted into channel C with certainty. On the other hand, if the
partial waves accumulate a relative phase difference of $\varphi=\pi$, the
photon is transmitted into mode D. Due to this single-photon interference
effect, the MZI distinguishes between certain qubit states: by placing two
dispersively interacting transmon qubits A and B in each arm of the
interferometer as shown in figure~\ref{fig:setup}, each partial wave picks up
an individual phase due to scattering from these distinct qubits. If the
qubits are initialized in a state spanned by the subspace
$\ket{\uparrow_\text{A},\uparrow_\text{B}}$,
$\ket{\downarrow_\text{A},\downarrow_\text{B}}$, the scattering phases on are
identical in both arms and the photon is transmitted into channel C. If on the
other hand, the qubits are in a state spanned by
$\ket{\uparrow_\text{A},\downarrow_\text{B}}$,
$\ket{\downarrow_\text{A},\uparrow_\text{B}}$ and if the scattering induces a
relative phase difference of $\varphi=\pi$, then the photon is transmitted
only into channel D. Hence, the interference of the partial waves allows for a
parity-selective transmission of the photon through the MZI, discriminating
the states $\ket{\uparrow_\text{A},\uparrow_\text{B}}$,
$\ket{\downarrow_\text{A},\downarrow_\text{B}}$ with even parity from the
states $\ket{\uparrow_\text{A},\downarrow_\text{B}}$,
$\ket{\downarrow_\text{A},\uparrow_\text{B}}$ with odd parity. As a
consequence of this parity-selective single-photon interference, the MZI can
be used to measure the qubit parity $P=\sigma_{z,\text{A}}
\sigma_{z,\text{B}}$.  Once the photon is registered in one of the detectors C
or D, the wave function collapses onto a state with definite parity, $P=1$
(even parity) or $P=-1$ (odd parity). Furthermore, such a parity measurement
is useful for entanglement preparation like other solid-state implementations
suggest \cite{Ruskov:2003zr, Beenakker:2004mz, Trauzettel:2006ly,
Williams:2008pd, Haack:2010fj} but here entanglement is accomplished in a
single-shot.

The protocol proceeds as follows: we initialize the qubits in the
superposition $\ket{+_\text{A}, +_\text{B}}$ with $\ket{+_j}=(\ket{\uparrow_j}
+ \ket{\downarrow_j})/\sqrt{2}$ for each qubit $j=\text{A},\text{B}$. This
state is---among other possibilities---a suitable choice and due to the
dispersive interaction $\propto \sigma^\pdag_{z,j} a^\dag_j a^\pdag_j$ each
partial wave of the photon introduces state-dependent scattering phases
$\varphi_\uparrow$ and $\varphi_\downarrow$ to the qubit: $( \ket{\uparrow_j}
+ \ket{\downarrow_j} )/\sqrt{2}  \mapsto 
(e^{i\varphi_{\uparrow}} \ket{\uparrow_j} + 
e^{i\varphi_\downarrow}\ket{\downarrow_j})/\sqrt{2}$ within each arm.
Crucially, we demand the phase difference to be $\varphi_\downarrow -
\varphi_\uparrow =\pi$ in order to make the MZI parity-discriminating. Note
that such a large difference between the scattering phases can only be
implemented within the strong coupling regime. After the photon is reflected
off the cavities, it carries information about the qubit state as well as about
the path it has taken. As this makes the states distinguishable we use the
second beamsplitter (II) to erase the which-path information of the photon. In
terms of the output Fock modes C and D the resulting state reads
$(\ket{\Phi^-} \ket{1_\text{C},0_\text{D}}  + \ket{\Psi^-}
\ket{0_\text{C},1_\text{D}})/\sqrt{2}$. Recasting the final state in terms of
the Bell states
\begin{subequations}
\begin{align}
 \ket{\Phi^-}	
 &= 	\frac{1}{\sqrt{2}} \left(\ket{\uparrow_\text{A},\uparrow_\text{B}} - \ket{\downarrow_\text{A}, \downarrow_\text{B}}\right) 
		& \text{with~}  P&=1, \label{eq:Bell-Phi} \\
 \ket{\Psi^-}	
 &= \frac{1}{\sqrt{2}} \left(\ket{\uparrow_\text{A}, \downarrow_\text{B}} - \ket{\downarrow_\text{A}, \uparrow_\text{B}}\right)
		& \text{with~}  P&=-1 \label{eq:Bell-Psi},
\end{align}
\end{subequations}
reveals the parity-selectivity of the MZI and shows furthermore that, due to
the missing which-path information, the parity measurement is not able to
distinguish in which arm the photon has scattered. Hence, the state projection
leaves the qubits in an entangled state no matter in which detector the photon
is registered. Taking the measurement outcome as starting point,  every other
entangled two-qubit state can be prepared by means of single qubit gates. In
this sense, the protocol accomplishes deterministic entanglement in a
single-shot.

In the following, we will look at this scheme on a more formal level that
allows to discuss imperfections.  The crucial part of the photon propagation
is the scattering process with the cavities and the qubits inside of these. In
order to describe the cavity-qubit subsystems, we assume for simplicity that
each qubit $j$ is coupled to a single cavity mode with frequency $\omega_{c}$.
The qubit states are separated by an energy splitting $\hbar\Delta$ which is
considered to be far detuned from the resonance frequencies of the cavities,
$\omega_{c} \gg \Delta$. In this regime, the light-matter coupling gives rise
to a qubit-state dependent renormalization of the cavity frequency---the
dispersive shift $\chi \sigma_{z,j}$. Accordingly, each cavity-qubit subsystem
is then described by the Hamiltonian
\begin{align}\label{eq:hj}
 H_j = \hbar \omega^\pdag_{c} a_j^\dagger a_j^\pdag +\frac{\hbar \Delta}{2} \sigma_{z,j}
		+ \hbar \chi \sigma^\pdag_{z,j} a_j^\dagger a_j^\pdag,
\end{align}
where $a^\pdag_j,a^\dagger_j$ are creation and annihilation operators of the
cavity modes obeying the canonical commutation relations
$[a_j,a^\dag_l]=\delta_{jl}$.  A photon $b_{\text{in},j}(k)$, incident to arm
$j$ of the MZI with wave number $k>0$ and frequency $\omega_k=c|k|$, induces a
qubit-state dependent phase shift after being scattered off the cavity. This
process is completely characterized by the reflection coefficient
\begin{align}\label{eq:reflection-coefficient}
 r(\omega_k;\sigma_{z,j}) = 
 \frac{i(\omega_{c} + \chi \sigma_{z,j} - \omega_k) -\kappa/2}
		{i(\omega_{c} + \chi \sigma_{z,j} - \omega_k) +\kappa/2}
\end{align}
which relates incoming and outgoing modes of the MZI by $b_{\text{out},j} =
r_j(\omega_k;\sigma_{z,j}) b_{\text{in},j}$, see appendix \ref{sec:tl} and
\cite{Walls:2008rt}. In (\ref{eq:reflection-coefficient}), $\kappa$ denotes
the spectral broadening of the cavities. Due to the occurrence of the
$\sigma_{z,j}$ terms, the qubit states $\ket{\uparrow_j}$ and
$\ket{\downarrow_j}$ accumulate the relative phase difference $\varphi= \arg[
r(\omega_k;\sigma_z=1)] - \arg[ r(\omega_k;\sigma_z=-1)]$ while the photon
passes the interferometer. To achieve maximal entanglement between the qubits,
it is crucial to generate a relative $\pi$ phase shift, i.e., we would like to
adjust the parameters of the device such that
\begin{align}\label{eq:pi-phase}
 \pi =\arg\left[ r(\omega_k; \sigma_{z,j}=1)\right] -\arg\left[ r(\omega_k; \sigma_{z,j}=-1)\right]
\end{align}
for both qubits A and B. Condition (\ref{eq:pi-phase}) can be fulfilled by 
tuning the photon frequency to be
\begin{align}\label{eq:sweet-spot}
  \Omega = \omega_{c} \pm \sqrt{\chi^2 - \kappa^2/4}.
\end{align}
Note that these frequency sweet spots do only exist in the strong coupling
regime where $2\chi \geq \kappa$. For convenience we will only consider one
solution in (\ref{eq:sweet-spot}) and omit the other one; this particular
choice will be of no importance for the following analysis as long as we
consistently stick to it.  Recombining the two arms of the interferometer, the
second beamsplitter acts as linear transformation upon the outgoing modes A, B
and defines the output modes C, D via
\begin{align}
 	\begin{pmatrix} c_{\text{out}} \\ d_{\text{out}} \end{pmatrix} 
 &= \frac{1}{\sqrt{2}} \begin{pmatrix}  1 &  e^{i\theta} \\ -1 &  e^{i\theta} \end{pmatrix} 
	\begin{pmatrix}  b_{\text{out,A}} \\   b_{\text{out,B}} \end{pmatrix}.
\end{align}
The phase shift $\theta = k(\ell_\text{B}-\ell_\text{A})$ keeps track of to
the individual path lengths $\ell_j$ the photon has to take in each arm of the
interferometer. In the microwave regime, it can be tuned in situ by phase
shifters, e.g., build from a pair of dc-SQUIDS \cite{naaman:16}.   As we will
see later, this parameter turns out to be useful to prevent disturbing
interferences that negatively affect the degree of entanglement.  Finally,
after passing II, the photon is absorbed by one of the detectors C and D
thereby projecting the transmon qubits onto an entangled state.

However, several erroneous mechanisms may spoil the production of full
entanglement and lead to limitations of our scheme: the shape of the photon
wave-packet, the fine tuning of cavity parameters, dissipative photon
propagation due to leaky cavities or imperfect circulators, and a finite
quantum efficiency of microwave photon counters. The first two points are
difficulties intrinsic to the entanglement protocol. The latter points arise
because of (extrinsic) technological limitations: If the circuit elements or
the circulators are leaky, there is a finite probability that the photon is
lost during propagation. Additionally, the photon counters  do not work at
maximal efficiency.  These effects combined lead to the situation that the
photon is not detected in each round of the protocol. The fight these
extrinsic limitations, we envision our protocol to be conditioned on the
detection of a photon in either detector C or D. As a result, the entanglement
generation is heralded.  Apart from potential dark counts which are
detrimental in for any remote entanglement protocol, the conditioning makes
sure that the state of the qubit is projected onto a pure state. The extrinsic
limitation thus only reduce the success rate, i.e., the rate of entanglement
generation, but not fidelity, i.e., the purity of the entangled states.  The
focus of the remaining discussion is thus the fidelity of the state obtained
given the intrinsic limitations mentioned above.

For a non-ideal setup, the photon measurement yields---depending on the
outcome---projected states $\ket{\Psi^\text{m}}$ and $\ket{\Phi^\text{m}}$
that generally deviate from a Bell state. Note however, as there is only a
single photon present, the state conditioned on the detection of the photon is
still a pure state. In order to quantify the degree of entanglement, we
determine the fidelity of the outcome after the measurement and the wanted
Bell state; the fidelity is a measure of distance between states in Hilbert
space and is defined as $F[\psi,\phi] = |\langle \psi |\phi \rangle|^2$ for
two arbitrary pure states $\ket{\psi}$ and $\ket{\phi}$, see
\cite{Nielsen:2010fk}.  Specifically, we are interested in the fidelities
$F[\Phi^\text{m},\Phi^-]$ or $F[\Psi^\text{m},\Psi^-]$, depending on the
measurement outcome. If these quantities take the value one, the projected
state is the sought-after Bell state. In the following, we demonstrate how
intrinsic errors affect the entanglement production and compute how the
fidelity is affected by a single-photon wave-packet with Lorentzian wave
profile as well as for non-identical cavities.

\begin{figure}[t]
  \centering
  \includegraphics[width=0.6\textwidth]{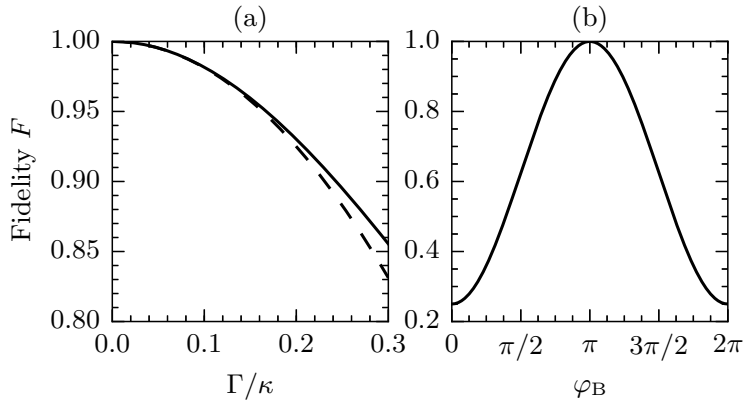} 
  \caption{Fidelities $F[\Psi^\text{m},\Psi^-]$ of the protocol for the two
  two intrinsic error mechanisms as discussed in the main text. In (a), the
  fidelity $F[\Psi^\text{m},\Psi^-]$ is shown as function of the spectral
  broadening in case a Lorentzian-shaped photon wave-packet passing the MZI
  (for $\chi=2\kappa$).  The solid line is obtained numerically whereas the
  dashed line corresponds to (\ref{eq:fidelity-gamma}); both results agree
  in the regime $\Gamma \ll \kappa$. In (b), the fidelity (\ref{eq:fidelity-phi}) 
  is shown as a function of the parameter $\varphi_\text{B}$, quantifying the 
  detuning of cavity B, while assuming that cavity A is perfectly tuned. 
  The ideal case corresponds to $\varphi_\text{B}=\pi$ with fidelity $F=1$.
  }\label{fig:fidelity}
\end{figure}

Generally, a single travelling photon is emitted as a wave-packet, i.e., a
superposition of various frequencies. In other words, it becomes impossible to
fulfil (\ref{eq:pi-phase}) for a generic photon state. For concreteness, we
assume that the single microwave photon is produced by the controlled decay
of a microwave resonator which means that the envelope function is a
Lorentzian wave-packet
\begin{align}
	f(k) &= \frac{ (c \Gamma)^{1/2}}{i(\omega_k-\Omega) - \Gamma/2}
\end{align}
with spectral broadening $\Gamma$ and the mean frequency tuned to $\Omega$.
Importantly, the relative weight factor $\eta$ between qubit states
$\ket{\uparrow_j}$ and $\ket{\downarrow_j}$, which is implied by the
dispersive interaction, can be represented as a coherent sum over all
frequency components in the wave-packet, see also appendix \ref{sec:fid}.
While the central frequency component of the photon wave-packet $\omega_k=
\Omega$ reveals a relative phase factor of $e^{i\pi}$, all other frequency
components induce relative phase factors deviating from this value. By
averaging coherently over all these contributions the resulting weight factor
$\eta$ has modulus less than one, $|\eta|<1$, and an average phase
$\varphi=\arg(\eta)\neq \pi$ which generally differs from $\pi$. In
particular, for the Lorentzian wave-packet the fidelity is a function of the
photon spectral width $\Gamma$. Assuming $2\chi>\kappa$ \footnote{In the case
$2\chi =\kappa $, the first non-vanishing contributions of (\ref
{eq:fidelity-gamma}) and
	(\ref{eq:fidelity-phi-approx}) appear to forth order in $\Gamma
/\kappa $ and $\delta \omega /\kappa $.}, $\theta=0$, and focusing to the
limit $\Gamma/\kappa \ll 1$ we find the fidelity
\begin{align}\label{eq:fidelity-gamma}
 	F[\Phi^{\rm m},\Phi^-]
 &\simeq 	1 - 2 \left[ 1- \left( \frac{\kappa}{ 2 \chi} \right)^2  \right] \,
 \left( \frac{\Gamma}{\kappa} \right)^2 ,
\end{align}
see figure~\ref{fig:fidelity}.
Equation~(\ref{eq:fidelity-gamma}) holds if the photon has been registered in
the detector C. For the reciprocal measurement outcome we find, due to
parity-selective interference amplitudes, $F[\Psi^\text{m},\Psi^-]=1$
irrespective of the line width, see appendix \ref{sec:fid} for details. As
$\Gamma/\kappa$ approaches zero, the fidelity (\ref{eq:fidelity-gamma})
becomes unity thereby achieving full, deterministic entanglement, i.e., for
any measurement outcome.

Moreover, in a realistic setup the cavities will always be fabricated slightly
differently. It is conceivable that  this can be partially remedied by tunable
cavities \cite{srinivasan:11,zhang:17}. However, in general, the frequency
sweet spots of cavity A and B differ from each other, $\Omega_\text{A}\neq
\Omega_\text{B}$, due to different cavity resonant frequencies and also
different cavity broadenings. This implies that we can, at best, tune the
(central) frequency of the photon to fulfil (\ref{eq:pi-phase}) on one side,
say $\Omega_\text{A}$. Then, according to (\ref{eq:sweet-spot}) the dispersive
interaction induces a relative weight factor $\eta_\text{A}=e^{i\pi}$ between
states $\ket{\uparrow_\text{A}}$ and $\ket{\downarrow_\text{A}}$ of qubit A.
Since the other cavity cannot fulfil (\ref{eq:sweet-spot}) at the same time,
it induces a distinct relative weight factor $\eta_\text{B} =
e^{i\varphi_\text{B}}\neq -1$ where $\varphi_\text{B}$ is a function of the
parameters $\omega_{c,\text{B}}$, $\chi_\text{B}$, and $\kappa_\text{B}$. By
choosing the phase of the interferometer $\theta$ properly, $\varphi_\text{B}$
remains the only parameter that cannot be controlled in situ.  Expressed as a
function of $\varphi_\text{B}$, the fidelity reads
\begin{align}\label{eq:fidelity-phi}
 F[\Phi^{\rm m},\Phi^-] = F[\Psi^\text{m},\Psi^-] &=  \frac{1}{8} \left[5-3\cos (\varphi_\text{B}) \right].
\end{align}
When the parameters of both cavities coincide $\varphi_\text{B}$ takes the 
value $\pi$ and we recover the ideal case as indicated by a unit fidelity. 

To be more specific, we want to discuss the case where cavities A and B only
differ by their resonance frequency $\delta \omega = \omega_{c,\text{B}} -
\omega_{c,\text{A}} \neq 0$ with all other parameters identical. Expanding
$\varphi_\text{B}$ in terms of the small parameter $\delta \omega/\kappa$ we
find $\varphi_\text{B} \approx \pi + 4 (1 - \kappa^2/4\chi^2 \,)^{1/2}
\,\delta\omega/\kappa$ for $\chi>\kappa/2$. Hence, the fidelity is
\begin{align} \label{eq:fidelity-phi-approx}
 F[\Phi^{\rm m},\Phi^-] &\simeq  1 - 3 \left[ 1- \left( \frac{\kappa}{ 2 \chi}
 \right)^2  \right] \, \left( \frac{\delta \omega }{\kappa } \right)^2 .
\end{align}
Equations~(\ref{eq:fidelity-gamma}) and (\ref{eq:fidelity-phi-approx}) suggest
to tune the cavities in both cases such that $\kappa$ is sufficiently large as
compared to the photon line width and the cavity detuning. However, since
(\ref{eq:sweet-spot}) sets an upper limit $\chi\geq\kappa/2$ for the magnitude
of $\kappa$, the optimal implementation is a tradeoff between $\kappa$ being
larger than the photon line width and the cavity detuning, but still smaller
than the dispersive shift.

In summary, we have proposed a novel method to entangle distant transmon
qubits by the interaction with a \emph{single} photon. Relying on the ability
to access the strong coupling regime as well as the discrete nature of the
Fock state microwave photons, our scheme represents a parity meter based upon
strong projective measurements. We have analysed the entanglement protocol
under non-ideal conditions by demonstrating the sensitivity of the entangeled
state produced to experimental imperfections such as a finite line width of
the photons as well as cavity imperfections.
Furthermore, we have argued that for an optimal implementation the magnitude
of the cavity broadenings have to be on an intermediate scale limited from
above by the light-matter coupling strength.  Our protocol, can be used to
realise remote entanglement of two transmon qubits, a key ingredient for
distributed quantum computing.  The entanglement generation is on-demand,
heralded, and (in principle) deterministic. Different from prior proposals, it
is completely immune to photon loss and only negatively affected by the dark
counts of the detectors.

\section*{Acknowledgements}
We acknowledge financial support from the Alexander von
Humboldt-Stiftung.

\appendix

\section{Transmission line coupling}\label{sec:tl}

An arm of the MZI is considered to be a one-dimensional superconducting transmission line guiding microwave photons from beamsplitter I to beamsplitter II. Accordingly, each arm ($j=\text{A},\text{B}$) of the MZI hosts a one-dimensional continuum of modes
\begin{align}
H_{\text{tl},j}= \int \frac{dk}{2\pi} \, \hbar \omega_k \, b^\dag_j(k) b^\pdag_j(k).
\end{align}
The mode operators $b^\dag_j(k)$ and $b^\pdag_j(k)$, obeying the commutation relations $[b_j^\pdag(k),b_l^\dag(k')]=2\pi \delta_{jl} \delta(k-k')$, create and annihilate photon states from the vacuum in arm $j$ of the MZI. The wave number $k$ of a photon is related to its frequency $\omega_k$ by the linear dispersion relation $\omega_k = c |k|$. In our setup each transmission line (arm of the MZI) is intercepted by a microwave resonator enclosing a transmon qubit. For simplicity both cavities are characterized by a single mode described by operators $a_j^\dag$ and $a_j^\pdag$. The transmon qubit inside a cavity is dispersively coupled to the cavity mode. Therefore we can describe each cavity-transmon subsystem by the Hamiltonian 
\begin{align}
 H_j = \hbar \omega^\pdag_{c} a_j^\dag a_j^\pdag +\frac{ \hbar\Delta}{2} \sigma_{z,j} 
		+ \hbar \chi \sigma^\pdag_{z,j} a_j^\dag a_j^\pdag
\end{align}
as introduced in \eqref{eq:hj}. Photons inside the cavity can leak into the
transmission line and vice versa. This process is described by the coupling
Hamiltonian
\begin{align}\label{eq:tl-cav-coupling}
 H_{\text{cp},j}= \hbar \sqrt{c \kappa_j} \int \frac{dk}{2\pi} \left[ b^\dag_j(k) a_{j}^\pdag + a_{j}^\dag b^\pdag_j(k) \right]
\end{align}
where $\kappa_j$ characterizes the degree of hybridization between the cavity and its environment. Then, photons from the transmission line, which enter the cavity, interact with the qubit inside the cavity and are re-emitted into the transmission line. This scattering process is captured by the total Hamiltonian
\begin{align}
H_{\text{tot},j} = H_j + H_{\text{tl},j} + H_{\text{cp},j}.
\end{align}
Far apart from the cavities for $|x|\rightarrow \infty$, photons in the transmission lines are considered to be freely propagating with respect to $H_{\text{tl},j}$. This suggests the definition of the time-dependent input and output fields
\begin{subequations}
\begin{align}
 b_{\text{in},j}(x,t)  &= \int \frac{dk}{2\pi} e^{-i [\omega_k (t+T)-kx]} \, b_{\text{in},j}(k), \label{eq:in-state} \\
 b_{\text{out},j}(x,t) &= \int \frac{dk}{2\pi} e^{-i [\omega_k (t-T)-kx]} \, b_{\text{out},j}(k), \label{eq:out-state}
\end{align}
\end{subequations}
where the operators $b_{\text{in},j}(k)= b_j(k)|_{t=-T}$ and $b_{\text{out},j}(k)= b_j(-k)|_{t=T}$ generate asymptotically free scattering states for $T\rightarrow \infty$. For simplicity we will omit the space coordinate in the following and consider only the point $x=0$ where the transmission line is connected to the cavity. As follows from standard input-output theory for cavities \cite{Gardiner:1985fk, Walls:2008rt}, the input and output fields in (\ref{eq:in-state}) and (\ref{eq:out-state}) obey the boundary equation
\begin{align}\label{eq:input-output-boundary}
	b_{\text{out},j}(t) - b_{\text{in},j}(t) &= \sqrt{c\kappa_j} \, a_j(t).
\end{align}
Here $a_j(t)$ is the time-dependent cavity operator which is subject to the Heisenberg equations of motion 
\begin{subequations}\label{eq:Heisenberg}
\begin{align}
	\dot{ b}_j(k) 		&= 	-i \omega_k b_j(k) - i \sqrt{c \kappa_j} a_j, \label{eq:Heisenberg1} \\
	\dot{a}_j 			&= 	-i \left( \omega_{c,j} + \chi \sigma_{z,j} \right) a_j 
							- i\sqrt{c \kappa_j} \int \frac{dk}{2\pi} b_j(k),  \label{eq:Heisenberg2}  \\
	\dot{\sigma}_{z,j}	&= 	0.
\end{align}
\end{subequations}
The formal solution of (\ref{eq:Heisenberg1})
\begin{align}
 b_j(k;t) &= e^{-i\omega_k (t+T)} b_{\text{in},j}(k) 
 		   + \sqrt{c\kappa_j} \int_{-T}^t dt^\prime e^{-i\omega_k (t-t^\prime)} a(t^\prime)
\end{align}
can be put into (\ref{eq:Heisenberg2}). As a result, one obtains the quantum Langevin equation
\begin{align}\label{eq:Langevin}
	\dot{a}_j = -i\left( \omega_{c,j} +\chi_j \sigma_{z,j}\right) a_j - \frac{\kappa_j}{2} a_j
				+\sqrt{c\kappa_j}\, b_{\text{in}}(t)
\end{align}
where the input field plays the role of an external driving. By using (\ref{eq:input-output-boundary}) and the quantum Langevin equation, one can derive the input-output relation
\begin{align}\label{eq:input-output}
	b_{\text{out},j}(k) = r_j(\omega_k;\sigma_{z,j}) \, b_{\text{in},j}(k)
\end{align}
where input and output fields are related to each other via the frequency-dependent reflection coefficient 
\begin{align}\label{eq:reflection-coefficient_a}
 r_j(\omega_k;\sigma_{z,j}) = 
 \frac{i(\omega_{c,j} + \chi_j \sigma_{z,j} - \omega_k) - \kappa_j/2}
	{i(\omega_{c,j} + \chi_j \sigma_{z,j} - \omega_k) + \kappa_j/2},
\end{align}
cf.\ (\ref{eq:reflection-coefficient}). Note that due to the lack of transmission into other channels than the reflection channel, the reflection coefficient is a complex number with modulus one, $|r_j|=1$. Furthermore, the complex phase of $r_j(\omega_k;\sigma_{z,j})$ is the qubit state-dependent scattering phase that a photon acquires after being reflected from one of the two cavities.

\section{Fidelities of the projected wave functions}\label{sec:fid}

In order to obtain a general expression for the fidelities, it has to be taken into account that photons usually are emitted as wave-packets. Therefore it is convenient to introduce wave-packet operators 
\begin{align}
	B^\dag_{j} = \int \frac{d k}{2\pi} \, f(k)\, b_j^\dag(k)
\end{align}
that create single photons with a certain wave profile $f(k)$ from the vacuum state in arm $j$ of the interferometer. To ensure that these wave-packets carry the intensity of one photon, the envelope function has to be normalized to one, $\int \frac{dk}{2\pi} |f(k)|^2 = 1$. Accordingly, the state of an incoming photon, which has been split by beamsplitter I, can be represented as 
\begin{align}
 \ket{\psi_\text{ph}} &= \frac{1}{\sqrt{2}} (B^\dag_\text{A} + B^\dag_\text{B}) \ket{0}. 
\end{align}

After the photon has scattered with the qubits inside the MZI, the two modes $b_{\text{out,A}}$ and $b_{\text{out,B}}$ of arms A and B are converted into the output modes
\begin{align}
	c_\text{out} 	&= \frac{1}{\sqrt{2}}\left(b_{\text{out,A}}+ e^{i\theta} b_{\text{out,B}} \right), \\
	d_\text{out}	&= \frac{1}{\sqrt{2}}\left(-b_{\text{out,A}}+ e^{i\theta} b_{\text{out,B}} \right)
\end{align}
by the second beamsplitter (II) to erase the which-path information of the photon. The phase shift $\theta = k(\ell_\text{B} - \ell_\text{A})$ arises due to the difference of the optical paths of the photon in the interferometer. Then, by detecting the photon in one of the output modes C or D, the total wave function $\ket{\psi}=\ket{\psi_\text{qb}} \otimes \ket{\psi_{\text{ph}}}$ is projected onto one of the states  
\begin{align}
	\ket{\Phi^\text{m}}	&= \frac{c_\text{out}(t) \ket{\psi}}{\langle \psi |c^\dag_\text{out} c^\pdag_\text{out} |\psi \rangle^\frac{1}{2}}, 
	& \text{or} &
 	&	\ket{\Psi^\text{m}}	&= \frac{d_\text{out}(t) \ket{\psi}}{\langle \psi |d^\dag_\text{out} d^\pdag_\text{out} |\psi \rangle^\frac{1}{2}}. \nonumber
\end{align}
In particular, the projected wave functions can be represented as
\begin{subequations}\label{eq:projected-states}
\begin{align}
\ket{\Phi^\text{m}}	
 &= \frac{1}{\sqrt{N_\Phi}}
	\sum_{\sigma,\sigma'} \left( f_{\sigma,\text{A}} + f_{\sigma',\text{B}} \right) \ket{\sigma_\text{A},\sigma'_\text{B}},
					\label{eq:Phi-m} \\
 	\ket{\Psi^\text{m}}	
 &= \frac{1}{\sqrt{N_\Psi}}
	\sum_{\sigma,\sigma'} \left( f_{\sigma,\text{A}} - f_{\sigma',\text{B}} \right) \ket{\sigma_\text{A},\sigma'_\text{B}}
					\label{eq:Psi-m}
\end{align}
\end{subequations}
with normalization constants $N_\Phi$ and $N_\Psi$. All information about the shape of the photon wave-packet as well as the cavity detuning, i.e., the differences between the two cavities is encoded into the linear coefficients which are given by
\begin{align}\label{eq:coefficients}
 f_{\sigma,j} =\int \frac{dk}{2\pi} \, r_j(\omega_k;\sigma_j) f(k)  e^{-i(\omega_k t - k x_j)}.
\end{align}
Here $x_j = \ell_j +\ell_0$ denotes the total distance, that has been taken by the photon from I through arm $j$ to the detectors; $\ell_j$ is the path length of arm $j$ and $\ell_0$ is the distance from II to the detectors. From the general expressions (\ref{eq:projected-states}) the ideal case is easily recovered: if the photon frequency is emitted exactly at the sweet spot $\omega_k=\Omega$ and if both cavities are identical, the projected states are equal to the favored Bell states $\ket{\Phi^\text{m}}=\ket{\Phi^-}$ and $\ket{\Psi^\text{m}} = \ket{\Psi^-}$.
 
However, if the conditions of the ideal case cannot be fulfilled, the projected state deviates from a maximally entangled state. This deviation can be quantified in terms of the fidelities of the measured states and the designated Bell-state of the ideal case,
\begin{align}
 	F[\Phi^\text{m},\Phi^-] 
 =& \frac{1}{2|N_\Phi|} |(f_{\uparrow,\text{A}} + f_{\uparrow, \text{B}}) - ( f_{\downarrow,\text{A}}+ f_{\downarrow,\text{B}}) |^2, \\
  	F[\Psi^\text{m},\Psi^-] 
 =& \frac{1}{2|N_\Psi|} |(f_{\uparrow,\text{A}} - f_{\downarrow, \text{B}}) - ( f_{\downarrow,\text{A}}- f_{\uparrow,\text{B}}) |^2.
\end{align}

So far, all expression have been presented in a general form including the effect of photon wave-packet propagation as well as the effect of detuned cavities. In the following each of these effects shall be discussed separately.

\subsection{Lorentzian wave packet}

To quantify the influence of a Lorentzian-shaped wave packet, the coefficients in (\ref{eq:coefficients}) have to be evaluated for a Lorentzian envelope function
\begin{align}
  f(k) &= \frac{(c \Gamma)^{1/2}}{i(\omega_k-\Omega) - \Gamma/2}.
\end{align}
Here it is assumed that the central frequency of the envelope function equals
the sweet spot frequency $\Omega= \omega_{c} \pm (\chi^2 - \kappa^2/4)^{1/2}$
in order to induce a $\pi$-phase shift, see (\ref{eq:sweet-spot}). Since both cavities are considered to be equal in this case, the linear coefficients are symmetric with respect to exchange of the cavities, $f_{\sigma,\text{A}}=f_{\sigma,\text{B}}$. Therefore, we will omit the index $j$ in the following. Then, the coefficients take two independent values $f_\uparrow$ and $f_\downarrow=\eta f_\uparrow$ where $\eta$ is a complex factor. Moreover, for a Lorentzian wave-packet the coefficients $f_\uparrow$, $f_\downarrow$, and $\eta$ can be evaluated explicitly. In the limit of large cavity damping as compared to the photon line width $\Gamma\ll \kappa$, the $\eta$-factor becomes time-independent and reads
\begin{align}\label{eq:eta-parameter}
	\eta(\Gamma)=
	\prod_{\nu=1}^2 \frac{i[\omega_{c}+(-1)^\nu\chi-\Omega] -\Gamma/2+(-1)^\nu \kappa/2}
					{i[\omega_{c}+(-1)^\nu\chi-\Omega] -\Gamma/2 + (-1)^{\nu-1} \kappa/2}.
\end{align}
For the sake of simplicity the optical path length difference has been set to zero here, i.e., $\theta=0$. Together with (\ref{eq:eta-parameter}) the fidelities evaluate to 
\begin{align}
 	F[\Phi^\text{m},\Phi^-]
 &=	\frac{1+|\eta(\Gamma)|^2 - 2|\eta(\Gamma)|\cos[\varphi(\Gamma)]}{3+3|\eta(\Gamma)|^2 + 2|\eta(\Gamma)|\cos[\varphi(\Gamma)]}, \label{eq:fidelity-gamma-phi}\\
 	F[\Psi^\text{m},\Psi^-] &= 1 \label{eq:fidelity-gamma-psi}
\end{align}
where $\varphi(\Gamma)=\arg[\eta(\Gamma)]$ denotes the frequency-averaged phase difference.

At first glance the result in (\ref{eq:fidelity-gamma-psi}) seems
surprising. In fact, the fidelity takes the value one because both cavities
have been assumed to be identical here. This implies that both states
$\ket{\uparrow_\text{A}, \uparrow_\text{B}} \otimes \ket{1_\text{A},
0_\text{B}}$ and $\ket{\uparrow_\text{A} \uparrow_\text{B}} \otimes
\ket{0_\text{A}, 1_\text{B}}$ obtain exactly the same scattering phase while
traversing the interferometer. Clearly, the same argument must also hold for
the pair of states $\ket{\downarrow_\text{A} \downarrow_\text{B}}\otimes
\ket{1_\text{A}, 0_\text{B}}$ and $\ket{\downarrow_\text{A}
\downarrow_\text{B}}\otimes \ket{0_\text{A}, 1_\text{B}}$. Since each of these
pairs have the same phase, they interfere destructively in output channel D.
In other words, due to parity-selective interference only states with odd
parity can be transmitted into channel D. Accordingly, by registering the
photon in channel D, the wave function is necessarily projected onto the Bell
state with odd parity---irrespective of the photon line width. Note that the
fidelity (\ref{eq:fidelity-gamma-psi}) will depart from one as soon as the
two cavities become unlike.

\subsection{Detuned cavities}

Microwave resonators, which are not identically fabricated, generally differ
in their resonance frequencies, $\omega_{c,\text{A}} \neq
\omega_{c,\text{B}}$, their qubit couplings $\chi_{\text{A}}\neq
\chi_{\text{B}}$, and their cavity broadenings $\kappa_\text{A} \neq
\kappa_\text{B}$. Consequently, for a photon, that is emitted at a single
frequency, it is impossible  to match the sweet spot frequency for both
cavities simultaneously, see \eqref{eq:sweet-spot}. Assuming that the photon is emitted at frequency $\Omega_\text{A}$, the scattering-induced phase difference is exactly $\pi$ as indicated by the corresponding $\eta$-factor for cavity A, $\eta_\text{A} = f_{\downarrow,\text{A}}/ f_{\uparrow,\text{A}}=e^{i\pi}$. However, since the photon frequency does not coincide with $\Omega_\text{B}$, the corresponding $\eta$-factor for cavity B is $\eta_\text{B}=f_{\downarrow,\text{B}}/ f_{\uparrow,\text{B}}=e^{i\varphi_\text{B}}$ with $\varphi_\text{B}\neq \pi$. In addition, arising due to the detuning of the cavities, the complex factor $\eta_\text{AB}=f_{\uparrow,\text{B}}/f_{\uparrow,\text{A}}\neq1$ can be defined  in analogy to $\eta_\text{A}$ and $\eta_\text{B}$. Since we neglect the influence of a finite line width of the photon here, all linear coefficients $f_{\sigma,j}$ and their corresponding $\eta$-factors are complex phase factors with modulus one. 

One might expect that $\eta_\text{AB}$ has an influence on the fidelity, but fortunately this phase shift can be compensated by appropriately adjusting the optical path lengths inside the MZI. By using (\ref{eq:coefficients}) we find
\begin{align}
 	\eta_\text{AB}	&=\frac{f_{\uparrow,\text{B}}}{f_{\uparrow,\text{A}}} 
 				= \frac{r_\text{B}(\Omega_{\text{A}};\uparrow)}{r_\text{A}(\Omega_{\text{A}};\uparrow)}
 				  \,e^{i\Omega_\text{A}(\ell_\text{B}-\ell_\text{A})/c}.
\end{align}
Hence, by properly tuning the angle $\theta$ of the interferometer, $\eta_\text{AB}=1$ can be achieved.

\end{document}